\documentclass[twocolumn,showpacs,preprintnumbers]{revtex4}
\usepackage{graphicx}
\usepackage{dcolumn}
\usepackage{bm}

\begin{document}                

\title{Interfacial domain formation during magnetization reversal in exchange biased CoO/Co bilayers}
\author{F. Radu\footnote{Permanent address:
Department of Experimental Physics, National Institute of Physics and Nuclear Engineering, P. O. BOX MG-6, 76900, Magurele, Romania.} , M. Etzkorn, R. Siebrecht, T. Schmitte, K. Westerholt, H. Zabel}
\affiliation{Institut f\"{u}r Experimentalphysik/Festk\"{o}rperphysik,
Ruhr-Universit\"{a}t Bochum, 44780 Bochum,  Germany}

%
\begin{abstract}                
We have carried out detailed experimental studies of the exchange
bias effect of a series of  CoO/Co(111) textured bilayers with
different Co layer thickness, using the magneto-optical Kerr
effect, SQUID magnetometry, polarized neutron reflectivity, x-ray
diffraction, and atomic force microscopy. All samples exhibit a
pronounced asymmetry of the magnetic hysteresis at the first
magnetization reversal as compared to the second reversal.
Polarized neutron reflectivity measurements show that the first
reversal occurs via nucleation and domain wall motion, while the
second reversal is characterized by  magnetization rotation.
Off-specular diffuse spin-flip scattering indicates the existence
of interfacial magnetic domains.  All samples feature a small
positive exchange bias just below the blocking temperature,
followed by a dominating negative exchange bias field with
decreasing temperature.

\end{abstract}
\pacs{75.60.Jk, 75.70.Cn,  61.12.Ha}

\maketitle

\section{Introduction}

The exchange bias phenomenon is associated with the interfacial
exchange coupling between ferromagnetic and antiferromagnetic spin
structures,  resulting in an unidirectional magnetic anisotropy
 that causes a shift of the hysteresis loop to negative
field values as one cools the system through the N\'{e}el temperature
of the antiferromagnet(AF) in a positive magnetic
field \cite{bean}. The exchange bias effect is essential for the
development of magneto-electronic switching devices (spin-valves)
and for random access magnetic storage units. For these
applications a predictable, robust, and tunable exchange bias
effect is required.

Extensive data have been collected on the exchange bias field
$H_{EB}$, and the coercivity field $H_c$, for a large number of
bilayer systems, which are reviewed in Ref. \cite{berk,nogu}. The
details of the EB effect depend crucially on the AF/F combination
chosen and on the structure and thickness of the films. However,
some characteristic features apply to most systems: (1) $H_{EB}$
and $H_c$ increase as the system is cooled in an applied magnetic
field below the blocking temperature $T_B \leq T_N$ of the AF
layer, where $T_N$ is the N\'{e}el temperature of the AF layer; (2)
the magnetization reversal can be different for the ascending and
descending part of the hysteresis loop \cite{radu}, as was first
pointed out in reference \cite{fitz}, and recently also observed
by Lee et al. \cite{lee} and Gierlings et al. \cite{gier}; (3)
thermal relaxation effects of $H_{EB}$ and $H_c$ indicate that a
stable magnetic state is reached only at very low temperatures
\cite{radurelax, good, geog}. Furthermore, a positive $H_{EB}$ has
been observed after cooling an AF/F system in very high magnetic
fields ~\cite{nogu1,hong}, and a perpendicular exchange bias was
realized by using a ferromagnetic layer with perpendicular
anisotropy \cite{maat}. Several theoretical models have been
developed for describing possible mechanisms of the EB effect,
including domain formation in the AF layer with domain walls
perpendicular to the AF/F interface \cite{malo}, creation of
uncompensated excess AF spins at the interface \cite{schul}, or
the formation of a domain wall in the AF layer parallel to the
interface \cite{maur,stil}. Another approach is the consideration
of diluted antiferromagnets in a field. In the work of Milt$\acute{e}$nyi et
al., Keller et al. \cite{milt,kell}, and Nowak et al.\cite{nowa}
the discussion about compensated versus uncompensated interfacial
spins is replaced by a discussion of net magnetic moments within
the antiferromagnetic layer\cite{hoff}. They show that perpendicular magnetic
domain walls in the AF layer can effectively be pinned at defect
sites, providing an enhanced exchange bias field. Recent overviews
of theoretical models on the exchange bias effect are provided by
Stamps \cite{stam} and by Kiwi \cite{kiwi}. However, so far there
is still not a satisfactory agreement between the different
theoretical approaches and experimental observations of the
macroscopic and microscopic response of exchange bias systems.

One step forward  in describing the exchange field behavior, as we
will demonstrate in the present report, could come from a detailed
analysis and understanding of the asymmetric behavior of the
magnetization reversal and relaxation processes of the
magnetization close to the coercive fields. To this end, we have
chosen to investigate the archetypal CoO/Co(111) AF/F exchange
bias system because it offers a number of advantages over similar
systems, which show also the EB effect. First, Co has very good
growth properties as a thin film and very flat interfaces can be
prepared by sputtering and molecular beam epitaxial techniques.
Second, the N\'{e}el temperature of the oxide is at a convenient
temperature of $291$~K. Third, CoO provides a large EB even for
very thin layers, which can easily be prepared by thermal
oxidation \cite{gruy}. Furthermore, CoO/Co bilayers exhibit
straightforward and pronounced EB properties, which qualifies them
as a model system for detailed investigations of the magnetization
reversal process~\cite{gruy,gier,milt,felcher}. This is the case
although the spin structure of the antiferromagnetic CoO is rather
complex. In CoO the spin-orbit interaction and therefore the
crystal anisotropy is strong. For bulk CoO the easy axis is
reported to be along the [117] directions \cite{Roth}.
However, in very thin layers because of missing neighbors the easy
axis has still to be determined. Training effects, which have been
observed for CoO/Co, add an additional difficulty. However,
training effects can also be useful for the understanding of the
metastability of interfaces, as we will show further below.

In the present paper we investigate the thickness dependence of
the exchange bias effect by using a wedge type CoO/Co(111) bilayer
with a Co thickness varying from 27~\AA~to 400~\AA. Previous
studies by Gruyters and Riegel \cite{gruy} have shown that the
shape of the hysteresis changes dramatically with the Co
thickness, indicating that aside from film orientation and
exchange coupling the film thickness plays an important role for
the type of domains formed and magnetization reversal observed.
Here we report extensive hysteresis measurements of CoO/Co(111)
bilayers, including a detailed analysis of the interfacial domain
state by investigating the specular and off-specular neutron spin
flip cross section at the coercive fields for descending and
ascending magnetic fields.

\section{Sample Preparation and Characterization}

\begin{figure}[ht]
\includegraphics[clip=true,keepaspectratio=true,width=1\linewidth]{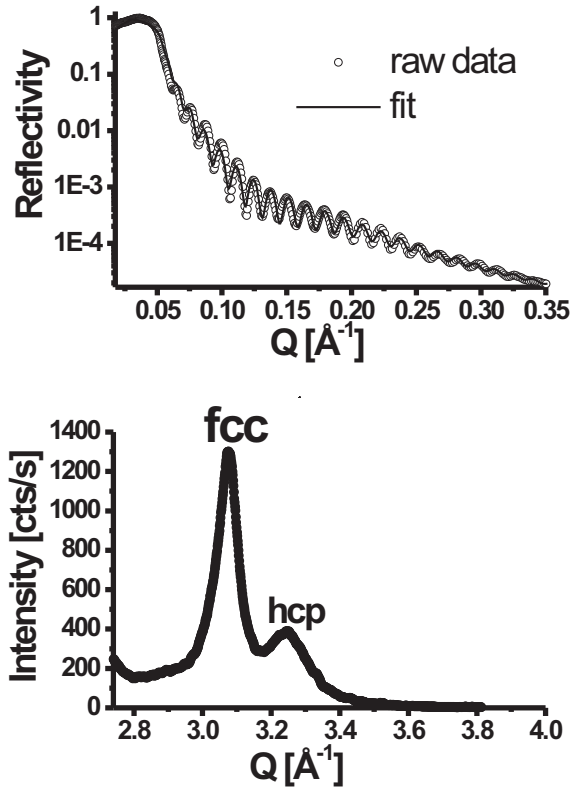}
 \caption{\label{xrays5}Top panel: Reflectivity measurement of sample s9 (CoO(25\AA)/Co(400\AA)).using
 synchrotron radiation  (open dots). The solid line shows a best fit to the data points.
Bottom panel: X-ray radial Bragg scan parallel to the surface
normal of CoO(25\AA)/Co(150\AA). The main peak at $Q=$3.08~\AA$^{-1}$
is close to the ideal fcc Co(111) position, indicating a majority
fcc phase. The second peak is from a minority tilted hcp component
with [10$\bar{1}$1] orientation. The inset shows a transverse scan
through the Co(111) peak. }
 \end{figure}

A series of nine samples were prepared by  rf-sputtering  in 99.99\% Ar
(samples s1-s9 in Table. \ref{table1}). The a-plane Al$_2$O$_3(11\bar{2}0)$
substrates were kept at $300^{\circ}$ C during the
Co deposition, which is the optimized growth
temperature as concerns mosaicity, structural coherence length,
and surface morphology, as determined in Ref. \cite{Morawe}.
The substrate for sample s9 was placed in the
middle of the sample holder while the other 8 substrates were
placed at equal distances away from each other.
\begin{figure}[ht]
\includegraphics[clip=true,keepaspectratio=true,width=1\linewidth]{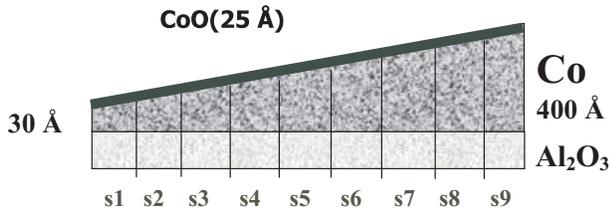}
\caption{\label{sample} The design of the wedge type CoO/Co(111)
sample grown on a sapphire substrate. Before growth, the Al$_2$O$_3$ substrate is
cut in 9 pieces and placed on the sample holder at equal distances away from each other.
The gradient growth rate provides samples s1 to s9 of different thickness. }
\end{figure}
Taking advantage
of the natural gradient in the sputtering rate, the samples
acquire an increasing thickness from 30~\AA~(s1) to 400~\AA~(s9)
(see Fig. \ref{sample} ). Subsequently, the samples were exposed to
air at room temperature which results in a 25~\AA~ thick CoO
layer on top of the Co film. The characterization of the surface
morphology by Atomic Force Microscopy (AFM) shows an exceptionally
low roughness of 1~\AA~ for all samples, which has also been
confirmed by x-ray reflectivity measurements performed at the W1.1
beam-line of the HASYLAB. As a representative example in Fig.
\ref{xrays5} the reflectivity is shown for the thickest sample
(s9) together with a best fit to the data points. The thickness
parameters obtained for all samples from the fits are listed in
Table \ref{table1}. High angle x-ray diffraction measurements
(Fig. \ref{xrays5}) show that the Co layer consists mainly of the
cubic fcc phase with the (111) growth direction yielding a strong
peak at $Q=3.08$~\AA$^{-1}$, and a minority hcp phase with a tilted
$(10\bar{1}1)$ growth responsible for the peak at $Q=3.25$~\AA$^{-1}$.
 The observed 'fcc' peak position is closer to the
ideal fcc position at $Q=3.071$~\AA$^{-1}$~than to the ideal hcp
position at $Q=3.106$~\AA$^{-1}$, which is in good agreement with
earlier measurements of Co films on sapphire substrates grown at
the same substrate temperature and similar
thicknesses\cite{Morawe}. The full width at half maximum (FWHM) of
the rocking curve of the pseudo Co(111) Bragg peak is 0.01 degrees
for an incoming wavelength of 1.393 \AA, confirming a very good
texture of the Co film.  Natural oxidation of the Co(111) films
leads to a preferential growth of CoO with (111) orientation. We
have studied, in addition, epitaxial Co(111) layers, which have
been thermally oxidized \textit{in-situ}. Synchrotron diffraction
studies of the in-plane CoO structure showed that [111] is the
surface normal of the 2.5 nm thick CoO film \cite{raduepi}. By
analogy we assume that for the highly textured CoO/Co bilayer
used for the present study, the same orientational relationship holds.
If the interface were ideal, this orientation would provide uncompensated
spins and therefore would exhibit an exchange bias effect. This
has been proven in the work of G\"{o}kemeijer et al.\cite{chien}, who
showed that for CoO/Co single crystalline films an exchange bias
occurs only in the (111) orientation but not in the (110) and
(100) orientations. The presence of uncompensated spins at the
interface is one of the key requirements for achieving a shift of
the hysteresis loop \cite{chien,milt,kell,nowa}, other important
factors are AF exchange constant and AF magneto-crystalline
anisotropy \cite{malo,stam}.

We first show magnetic characterizations  of the Co films in the
unbiased state at room temperature by taking hysteresis loops with
the magneto-optical Kerr effect (MOKE). Several hysteresis loops
were taken with the field parallel to the film plane but with
different azimuth angles of the sample. A plot of the ratio
between the remanent magnetization and saturation magnetization
$M_{re}/M_{sat}$, versus the rotation angle $\phi$, reveals a
two-fold in-plane anisotropy, which is observed for all samples
and which is induced by the sapphire substrate. In addition a
small contribution from a four-fold anisotropy can be recognized
by closer inspection. These results are in complete agreement with
earlier hysteresis measurements of Co(111) films on Al$_2$O$_3
(11\bar{2}0)$ substrates \cite{Metoki}. A typical example for an
unbiased hysteresis loop recorded at 270~K is plotted in Fig.
\ref{5hys} for sample s4. It is characterized by a completely
symmetric shape and a small coercive field of 60~Oe.  Additional
temperature dependent studies of the coercive field showed that
the N\'{e}el temperature of a 25 \AA ~thick CoO layer is about the
same as in the bulk ($T_N=$291~K). In fact, it might even be
slightly higher than in the bulk as shown in Ref.\cite{zaag}.
\begin{figure}[ht]

\includegraphics[clip=true,keepaspectratio=true,width=1\linewidth]{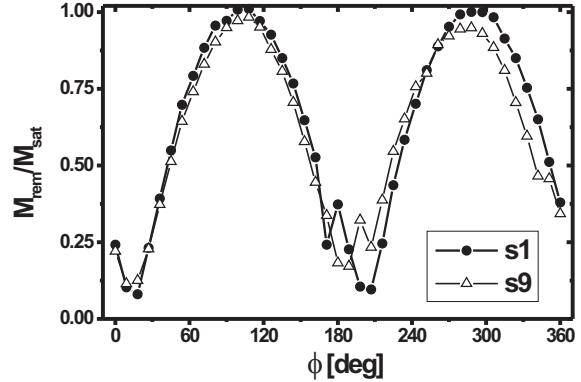}
\caption{\label{anys} Remanent magnetization divided by  the
saturation magnetization versus the rotation angle  of the sample
CoO(25\AA)/Co(26 \AA) (full circles) and  CoO(25\AA)/Co(400 \AA) (
open triangles) showing. The two fold anisotropy was observed for
all samples.}
\end{figure}

\section{Temperature dependence of the exchange bias
and coercive fields}

\begin{figure}[ht]

\includegraphics[clip=true,keepaspectratio=true,width=1\linewidth]{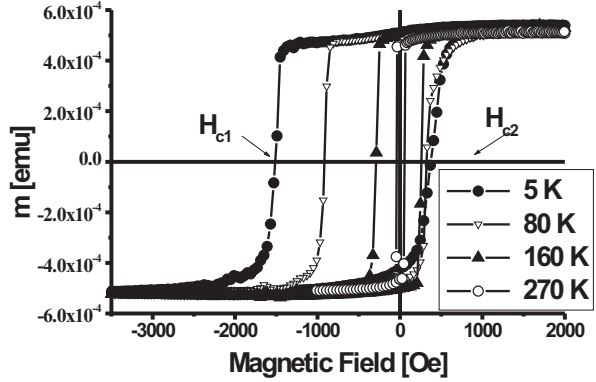}
\caption{\label{5hys} Hysteresis loops for  different temperatures
of sample s4 (CoO(25\AA)/Co(120\AA)). For each hysteresis the
bilayer was cooled in a field of +2000 Oe from 320K to the
respective temperature. The hysteresis curves were taken after
cooling and before training.}
\end{figure}

The temperature dependence of the hysteresis loops for samples
s1-s9 was measured by a  superconducting quantum interference
device (SQUID) magnetometer. The exchange bias field $H_{EB}$, the
coercive fields $H_{c1}$ and $H_{c2}$, and the half width of the
total coercive field $H_c=|(H_{c1}-H_{c2})/2|$) were extracted
from the hysteresis loops measured at temperatures between $310$
and $5~K$. The samples were cooled  from above the N\'{e}el
temperature to below different target temperatures in a magnetic
field of $H_{FC}$ =  +2000~Oe applied closely along the easy axis
of the sample magnetization. The procedure was repeated for each
target temperature. In Fig. \ref{5hys} typical hysteresis loops
are shown for the sample s4 with a Co thickness of $87$~\AA~ and
for the target temperatures 270~K, 160~K, 80~K, and 5~K. The
hysteresis curves were taken directly after cooling and thus  before
training, unless otherwise stated. The hysteresis loops show the
following typical and general features: $H_{c1}$ increases
strongly with decreasing temperature, while $H_{c2}$ remains
almost constant at a field value of about 300 Oe. The slope of the
hysteresis loops at $H_{c1}$ is steeper than at $H_{c2}$ on the
return path.

\begin{figure}[ht]
\includegraphics[clip=true,keepaspectratio=true,width=1\linewidth]{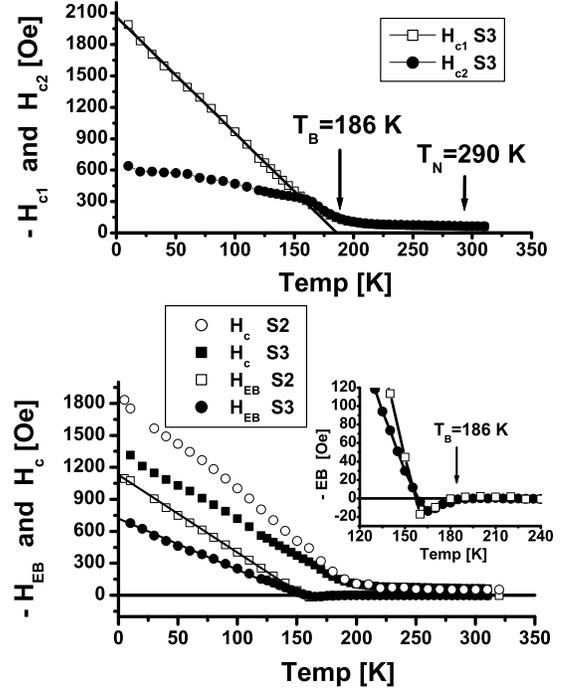}
\caption{\label{poseb}Top panel: Coercive fields $H_{c1}$ and $H_{c2}$ of sample s3.
The line is a fit to the linear region of $H_{c2}$. It intersects the
abscissa at the blocking temperature $T_B$ = 186 K.\\
 Bottom panel: temperature dependence of the
coercive field and the exchange bias field for samples s2 and s3.
In the inset the positive exchange bias occurring close to the
blocking temperature is shown. The lines are guides to the eye. }
\end{figure}

Fig. \ref{poseb} summarizes the analysis of the temperature
dependence of $H_{c1}$, $H_{c2}$, $H_{c}$, and $H_{EB}$ for the
samples s2 and s3.   The top panel reproduces the coercive field
of the first reversal $H_{c1}$ and of the second reversal $H_{c2}$
for the samples s3. Both coercive fields start to slightly
increase just below the N\'{e}el temperature with the same rate of
0.21 Oe/K.

\begin{figure}[ht]

\includegraphics[clip=true,keepaspectratio=true,width=1\linewidth]{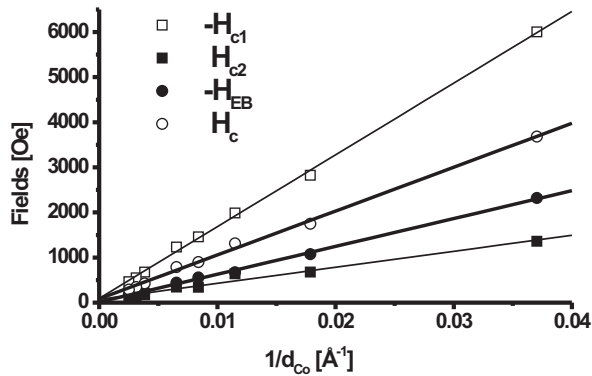}
\caption{\label{ddep}$-H_{c1}$ (open squares), $H_{c2}$ (full squares),
 $-H_{EB}$ (full circles), $H_{c}$ (open circles)
 are plotted as a function of the Co thickness. The samples
 were cooled down through the  N\'{e}el temperature of CoO to
 10 K in an applied magnetic field of +2000 Oe. The lines are
 linear fits to the data points.}
\end{figure}

 At the blocking
temperature of about $T_B \approx$ 186 K, the slope increases
drastically. Below $T_B$ a bifurcation for the temperature
dependence of $H_{c1}$ and $H_{c2}$ develops. While $H_{c1}$
keeps rising with a rate of 11.1 Oe/K, $H_{c2}$ levels off and
reaches saturation at the lowest temperature. Thus, there are
three distinguishable temperature regions: (1) from $T_N$ to $T_B$
the coercive fields are equal and increase slowly; (2) close to
$T_B$ the slopes increase drastically and $H_{c1}$ is slightly
smaller than $H_{c2}$; (3) below $T_B$ both coercive fields
develop linearly but with different slopes such that the absolute
value of $H_{c1}$ is bigger than $H_{c2}$. Only in this last
region a strong negative EB is observed. The different temperature
dependencies of both coercive fields are consistent with the
different magnetization reversal processes, which are recognized
via polarized neutron reflectivity measurements at these points to
be discussed further below. Previous studies of CoO/Co thin film
systems have shown an inverse proportionality between the exchange
bias field and the Co layer thickness \cite{gruy}. Our experiments
completely confirm these results and extend the linearity down to
a thickness of  27~\AA~ (see  Table. \ref{table1} and Fig.
\ref{ddep}).

In Fig. \ref{poseb} the exchange bias and coercive fields for
samples s2 and s3 are shown. At $T_B$=186~K~ the $H_{EB}$ becomes
positive and reaches a maximum of $H_{EB}$ = +20~Oe. The temperature
region of the positive EB is shown on an enlarged scale in the
insert of Fig. \ref{poseb}. This surprising feature has been
observed for all our samples. After changing sign, $H_{EB}$
decreases steadily as the temperature is lowered. Because of the
sign change we define the blocking temperature $T_B$ as the
temperature where $H_{EB}$ first deviates from zero. This
definition coincides with a linear extrapolation of $H_{c1}$ to
zero, shown by the solid line in the bottom panel.

\begin{figure}[ht]
\includegraphics[clip=true,keepaspectratio=true,width=1\linewidth]{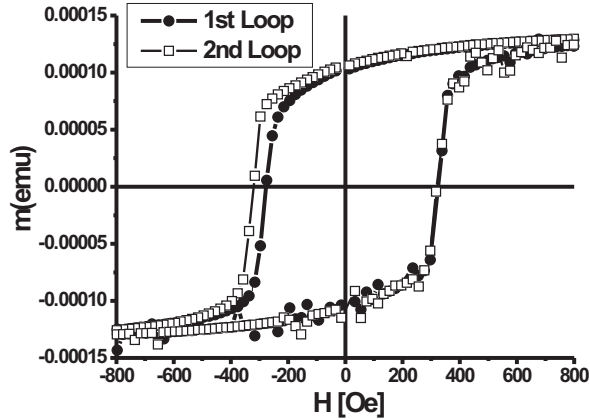}
\caption{\label{pextr} First (solid circles) and second hysteresis
(open squares) loop taken at 175 K. For the second hysteresis
loop  H$_{c1}$ decreased while H$_{c2}$ remained constant, implying a
 reduction of the positive exchange bias after training.}
\end{figure}

The linear dependence of $H_{c}$ and $H_{EB}$ on the temperature
is in good agreement with reports from other CoO/Co bilayers
\cite{schl,gruy, milt}. A positive $H_{EB}$ close to $T_B$ is seen
in our bilayers for the first time \cite{webeb}. Positive exchange
bias fields have been reported in the literature, but only for
high cooling fields, when the external field exceeds the
interfacial coupling, breaking the parallel alignment between the
F and AF layer~\cite{nogu1,hong}. Recently, three further papers
have reported about the observation of a positive exchange bias
effect \cite{gred,hell,full}. In contrast to those papers we
observe a weakening of the positive exchange bias after training.
This is shown in Fig. \ref{pextr}.

The positive exchange bias is considered as a proof for an
antiparallel alignment of the spins at the AF/F interface in
moderate cooling fields \cite{nogu1}. For parallel alignment of
the magnetic moments in the F and AF layers the bias field would
not change sign even in high fields. Therefore, the change of sign
from negative to positive exchange bias as a function of applied
field hints to antiferromagnet interfacial coupling requiring a
superexchange mechanism working at the F/AF
interface\cite{parker}.

\begin{figure}[ht]
\includegraphics[clip=true,keepaspectratio=true,width=1\linewidth]{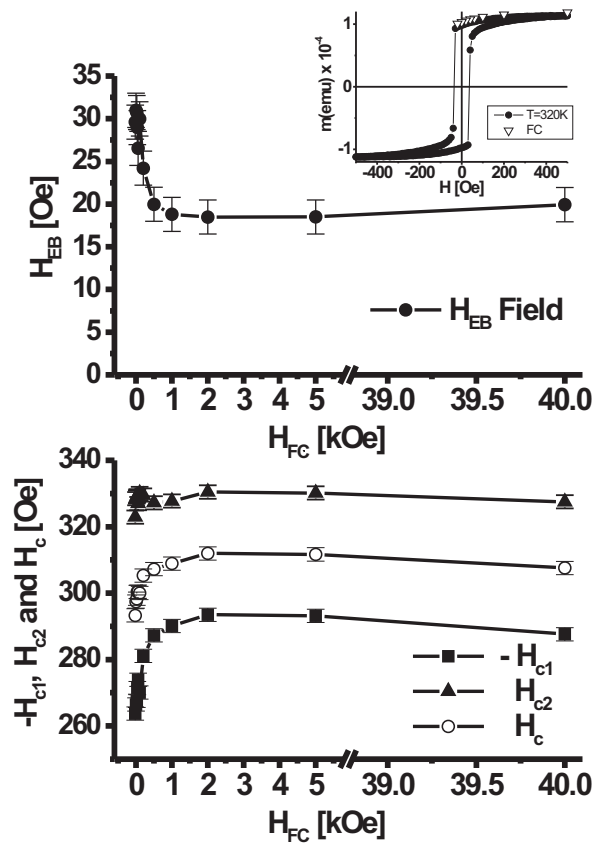}
\caption{\label{FCdep}Top: The field cooling dependence of the
positive exchange bias for sample s2 (CoO(25\AA)/Co(56\AA)). The
bilayer was cooled in different fields from above the N\'{e}el
temperature of CoO to 175 K, where hysteresis loops were measured.
The inset shows a hysteresis loop at T=320~K (full dots), together
with the cooling field values (down triangles) as reference of the
initial state of the sample. Bottom: The dependence of the
coercive field values ($-H_{c1}$, $H_{c2}$ and
$H_{c}=(-H_{c1}+H_{c2})/2$) as a function of the applied cooling
field. The lines are guides to the eyes. }
\end{figure}

In order get more insight into the origin of the positive exchange
bias, we have measured its field cooling dependence shown in Fig.
\ref{FCdep}. The sample s2(see  Table. \ref{table1})  was cooled
down from T=320 K to 175 K in different applied fields, ranging
from -20~Oe to 40~kOe (see the inset of (Fig. \ref{FCdep}). First
observation is that the exchange bias is positive for all cooling
fields. Second the bias field is constant and about +20~Oe between
1~kOe and 40~kOe, but below 1~kOe the bias field increases
dramatically to twice its former value. This increase is solely
due to a shift of $H_{c1}$ to smaller values with decreasing
cooling field, while $H_{c2}$ remains constant. We also note that
the increasing EB field correlates with a decreasing remanence at
320~K as the cooling field is lowered. Thus for small cooling
fields the initial magnetic state of the sample is not fully
saturated. At these small cooling fields ($<$~1000~Oe) we observe
a surprisingly strong increase of the EB field.

This behavior can be explained assuming two contributions to the
interfacial exchange coupling. Normally the interface exchange
coupling is ferromagnetic between CoO and Co as already mentioned
above, being the result of a direct exchange of Co monolayers on
either side of the ferro- and antiferromagnetic
layer\cite{parker}. For a positive exchange bias effect we have to
assume that some parts at the interface are, however, antiparallel
coupled. Then, an antiparallel alignment may be caused by a
superexchange type of mechanism mediated by an oxygen monolayer at
the interface. It is reasonable to assume that due to roughness or
thickness fluctuations, small fractions of the CoO film may be
oxygen terminated instead of metal terminated at the interface.

Next we discuss the training effect in the region below the
blocking temperature \cite{felcher,radu}, which is different from
the training for the positive exchange bias Fig.\ref{pextr}. After
closing a complete loop the hysteresis becomes symmetric and
assumes an S-like shape. The $H_{c1}$ field for all subsequent
loops and the EB-field is smaller than for the first reversal.
Thus the trained hysteresis has a different shape and a smaller EB
field than the hysteresis of the virgin sample. Obviously the
first reversal is dominated by a different process than the
following reversals, transforming the sample from a unique and
irreversible state to a different state with reproducible and
symmetric branches of the descending and
ascending parts of the hysteresis loop. When the sample is first
cooled in a positive field, the Co layer consists of a single
ferromagnetic domain. Upon cooling below the blocking temperature,
the orientation of antiferromagnetic domains in CoO is affected by
the exchange coupling at the AF/F interface. To optimize the
interfacial energy and to avoid spin frustration it is favorable
 \begin{table}[ht]
 \caption{\label{table1} Values for in-plane coercive and
 exchange bias
 fields and for Co layer thicknesses of a wedge type Co/CoO sample are listed.
 The thickness of the CoO layer is constant and about 25 \AA.
 The coercive fields  $H_{c1}$ and $H_{c2}$ and the exchange bias field $H_{EB}$
 are for samples which were cooled  in a field of 2000 Oe from
 above the
 N\'{e}el temperature(~290 K) to 10 K.}
 \begin{tabular}{|l|l|l|l|l|l|} 
\tableline
Sample index&$d_{Co}$[\AA] & $H_{c1}$[Oe] & $H_{c2}$[Oe] & $H_{EB}$[Oe]&$H_{c}$\\
\tableline
s1&27&-6004&1362&-2321&3683\\
s2&56&-2825&677&-1073&1751\\
s3&87&-1990&640&-675&1315\\
s4&119&-1458&343&-557&901\\
s5&153&-1233&346&-443&789\\
s7&260&-678&176&-251&427\\
s8&320&-549&144&-202&346\\
s9&398&-464&115&-174&290\\
\tableline
 \end{tabular}
 \end{table}
for the CoO film to develop a multi-domain state. However, after
cooling in the saturation field of the ferromagnetic layer, the
frustrated spins at the interface may very well be preserved,
which would yield a CoO layer in a single domain state. The lower
the interface roughness, the higher will be the probability for a
single domain state. The first reversal at $H_{c1}$ changes the
magnetic state of the interface, leaving behind a more disordered
spin structure.  A more precise analysis of the processes which
occur at the CoO/Co interface is possible with polarized neutron
reflectivity measurements, to be discussed next.

\section{Polarized Neutron Reflectometry}

In order to  gain more insight into the reversal mechanism in
decreasing and increasing magnetic fields, we have performed
polarized neutron reflectivity (PNR) measurements on  Co/CoO
bilayers, using the angle dispersive neutron reflectometer ADAM at
the Institut Laue-Langevin, Grenoble~\cite{Adam} with a fixed
wavelength of 4.41 \AA. Preliminary results are presented in Ref.
\cite{radu}. The sample has been optimized for the needs of
neutron reflectivity and consists of a CoO(25\AA)/Co(200\AA)
bilayer grown under the same conditions as explained above, but  on a
Ti(2000 \AA)/Cu(1000 \AA)/Al$_2$O$_3$ neutron resonator template.
This configuration of the sample allows to observe with high
sensitivity  the ferromagnetic domains and the spin misalignment
at the interfaces\cite{radu2,radu3,ir}.

The method we used to study the magnetization reversal by PNR
includes the following concepts: (1) the dependence of  spin
asymmetry measured at one specific value of the wave vector
transfer depends only on the cosine of the angle ($\theta$)
between the magnetic induction in the magnetic layer and the
neutron polarization direction; (b) the spin-flip reflectivity is
 proportional to $\sin^2(\theta)$; (c) the off-specular scattering shows the existence
of magnetic domains within the magnetic layer; (d) the enhancement
of the neutron density achieved at the magnetic interface, using
the neutron resonator, increases the off-specular scattering from
the interfacial magnetic domain walls.

\begin{figure}[ht]
\includegraphics[clip=true,keepaspectratio=true,width=1\linewidth]{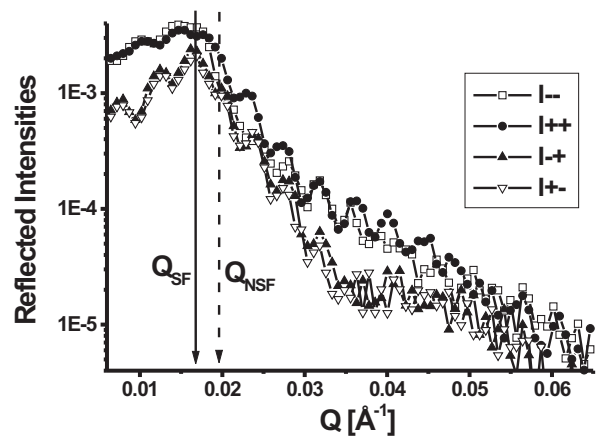}
\caption{\label{refl4} Polarized neutron reflectivity measurements
of a CoO/Co/Ti/Cu/Al$_2$O$_3$ resonator sample taken at room
temperature in a magnetic field of about 60 Oe (=H$_{c1}$).
The arrows for $Q_{NSF}$
and $Q_{SF}$ denote the points where non-spin flip and spin flip
intensities were recorded at low temperature after field cooling.
}
\end{figure}

In order to find the optimized scattering vectors for measuring
neutron hysteresis loops, we have first recorded a complete set of
polarized neutron reflectivities for all four cross-sections. The
raw data without  experimental corrections  (spin flip efficiency
and footprint) is shown in Fig. \ref{refl4}. The measurements were
carried out at room temperature and close to the coercive field.
Since at room temperature without exchange bias, reversal is
dominated by coherent rotation, the resonant peaks in the
spin-flip reflectivities are strong and well defined. We used this
information to choose the proper wavevector ($Q_{SF}$=0.016
$\AA^{-1}$) for recording the spin-flip intensity as a function of the
applied magnetic field at low temperatures. Similar, for the
non-spin flip intensities we have fixed the scattering vector just
above the critical edge for total external reflection of the
unpolarized beam ($Q_{NSF}$=0.019 $\AA^{-1}$). $SF$ and $NSF$
refer to spin flip and non-spin flip scattering, respectively.  We
have chosen this $Q_{NSF}$ value, because it provides the highest
contrast for non-spin flip reflectivity from the magnetic layer
combined with high intensity, whereas $Q_{SF}$ provides the
maximum intensity.

The magnetic hysteresis loop can be measured by PNR using the Spin
Asymmetry (SA) defined as:

\begin{equation}
SA(k_z,\theta,{\bf B} )=\frac{R^+ -R^-}{R^+ + R^-}=\frac{R^{++} -
R^{--}}{R^{++} + R^{+-} + R^{-+} + R^{--}}, \label{SA1}
\end{equation}

where $R^{++}$, $R^{+-}$, $R^{-+}$, and $R^{--}$ are the four
reflectivities from a magnetic multilayer system, $R^+=R^{++} +
R^{+-}$,  $R^-=R^{--} + R^{-+}$, and $\theta$ is the angle between
the  neutron polarization axis and the direction of the magnetic
induction ${\bf B}={\bf H}+4 \pi {\bf M}$, with ${\bf M}$  being
the magnetization of the ferromagnetic layer. $k_z$ is the
perpendicular component of the incoming wave vector. The spin
asymmetry, depends obviously on the optical parameters  and the
thickness of the layers. They have been omitted in the variable
list because they are constant.

For a  single magnetic layer embedded in a non-magnetic matrix the
SA is expressed as:
\begin{equation}
SA(k_z,\theta,{\bf B})=SA(k_z, 0, |{\bf B}|) cos(\theta)\approx
B_x/|{\bf B}|, \label{SA2}
\end{equation}
where $B_x=|{\bf B}| cos(\theta)$ is the component of the magnetic
induction parallel to the polarization axes. For a simple case we
derive this expression in the appendix.

The spin-flip reflectivities are:
\begin{equation}
 R^{\pm\mp} \approx sin(\theta)^2 \approx  B_y^2/|{\bf B}|^2,
\label{SF1}
\end{equation}
where $B_y=|{\bf B}| sin(\theta)$ is the perpendicular
(to the neutron polarization direction)
component of the magnetic induction.

The equations \ref{SA2} and \ref{SF1} are directly related to  the
perpendicular and parallel components of the magnetic induction of
a ferromagnetic layer. Moreover, through the spin flip scattering
one can decide whether the magnetization reversal is dominated by
rotation or domain wall movement.

\begin{figure}[ht]

\includegraphics[clip=true,keepaspectratio=true,width=1.1\linewidth]{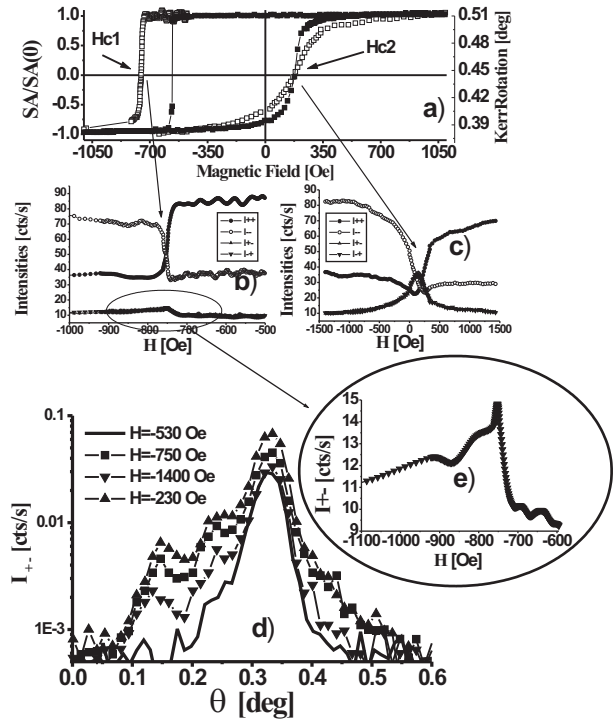}
\caption{\label{nhl1}   (a) MOKE hysteresis loop (full squares)
and neutron hysteresis loop (open squares) of a CoO/Co bilayer
after field cooling to 50K (MOKE) and to 10 K (neutrons) in an
external field of +2000 Oe.  (b) and (c) The specular non
spin-flip intensities $I_{ ++}$ and $I_{--}$ and the specular
spin-flip intensities $I_{+-}$ and $I_{-+}$ are plotted as a
function of external magnetic field. The intensities are measured
at special scattering vector values of the reflectivity curves
(see text); (d) Off-specular diffuse spin-flip scattering taken
before magnetization reversal at -530 Oe (line), at  $H_{c1}$=-750
Oe (full squares), in saturation at -1400 Oe (down full triangles)
and before the second magnetization reversal at -230 Oe (up full triangles); e) A zoom in from panel d) showing the $I_{+-}$ intensity at the first magnetization reversal.}

\end{figure}

The magnetic hysteresis $(SA(Q_{NSF},\theta,{\bf
B})/SA(Q_{NSF},0,|{\bf B_{sat}}|)=SA/SA(0)= cos(\theta)= B_x/|{\bf
B}|)$ loop  measured by PNR is shown in Fig. \ref{nhl1} together
with the magnetic hysteresis obtained by MOKE from the same
sample. The field dependencies of the non-spin flip (NSF)
reflectivities, from which the magnetic hysteresis is derived, are
shown in panels (b) and (c). The non-spin flip reflectivities were
measured at a wave vector transfer  $Q$ corresponding to the
inflection point of the non-polarized neutron reflectivity (near
the critical edge for total external reflection) and by sweeping
the magnetic field in the usual manner. Note that for technical
reasons the MOKE hysteresis curve was taken at 50 K while the
neutron measurements were performed at 10 K. Both curves show the
same asymmetric shape at $H_{c1}$ and $H_{c2}$ as discussed
before. The different $H_{c1}$ values for the MOKE and PNR
measurements are solely due to the different sample temperatures.
Nonetheless, we find very good agreement between both loops during
the first reversal. On the return path through $H_{c2}$ the
agreement is by far not as good as one would expect if no specular
intensity loss were occurring due to off-specular diffuse
scattering. On the other hand, this discrepancy between MOKE and
PNR is an indication for the presence of magnetic domains in the
ferromagnetic layer on the return path. This will be discussed in
more detail in the appendix (section IV.2).

In addition to the standard magnetization curves, PNR is capable
of distinguishing between different magnetization reversal
processes and to provide information on magnetic domains by
analyzing the specular and the off-specular spin flip (SF)
intensities $I^{+-}$ and $I^{-+}$. Specular SF scattering is
sensitive to magnetization components in the sample plane, which
are perpendicular to the applied field direction. The off-specular
SF signal reveals the presence of magnetic domains~\cite{toper}.

We first discuss the specular spin flip intensities  shown by
triangles in panels (b) and (c) of Fig. \ref{nhl1}. They were
measured at the scattering vector $Q$ corresponding to the resonance
peak near the critical edge.   The magnetization reversal at
$H_{c2}$ exhibits strong spin-flip intensities $I^{+-}$ and
$I^{-+}$. This is always observed for rounded or 'trained'
hysteresis loops and is characteristic for a magnetization
reversal via domain rotation \cite{dr}. Magnetization reversal by rotation
provides a large magnetization component perpendicular to the
field or polarization axis, giving rise to neutron spin-flip
process. Vice versa, the rather low spin-flip intensities, which
are observed during the first magnetization reversal in the virgin
state at $H_{c1}$ are indicative of pure 180$^\circ$ domain wall movement. The
step like intensity change at $H_{c1}$ =-750~Oe~ is followed by a
steady decrease as the system approaches saturation (see Fig. \ref{nhl1}e) . The slightly enhanced specular spin-flip scattering takes place in a very narrow field range of
not more than 15 Oe [FWHM=9 Oe]. From these
features it is obvious that domain wall nucleation and propagation
let the magnetic  spins at the interface canted away from the applied field direction.


Next we discuss the off-specular spin flip scattering close to
$H_{c1}$, which is reproduced in Fig. \ref{nhl1}d.   The
off-specular diffuse intensities taken at $H_{c1}$ =~-750~Oe, at
$H$=~ -1400~Oe, and upon return to $H$=~-230~Oe are rather strong.
Their asymmetric shape is attributed to the specific wave-vector
where they have been measured\cite{radu4}. Note that the off-specular
intensity appears only after the magnetization reversal at
$H_{c1}$ has taken place for the first time and is negligible
before. The solid line in Fig. \ref{nhl1}d is the SF scattering
taking in a descending field at -530~Oe (before first
magnetization reversal ), representing the instrumental
resolution. After reversal the diffuse scattering is large, then
decreases towards saturation in negative fields and increases
again on the way back towards H$_{c2}$. The diffuse spin flip
scattering reveals the existence of magnetic domains walls\cite{radu4}. We
argue that those domain walls are located at the interface between
the ferromagnetic and antiferromagnetic layer and that they have
to be distinguished from ferromagnetic domains in the Co layer. If
the diffuse scattering were originating from domain walls inside
of the Co-layer, then the off-specular scattering should vanish in
saturation, which is clearly not the case here. The creation of
interfacial domains is crucial for the change of the reversal
character from domain wall motion at $H_{c1}$ to domain rotation
at $H_{c2}$. The breakdown of the ferromagnetic film in domains
along the trained hysteresis has also been observed by Velthuis et
al. \cite{felcher}. Here we show that the breakdown occurs already
during the very first reversal and does not heal anymore even in
saturation. Opposite to Velthuis et al. \cite{felcher} we argue
that the observed domain formation is the reason for the trained
hysteresis curve, instead of training being the cause for a
partial breakdown of the alignment of the ferromagnetic domains.

The  PNR results lead us to the following interpretation of the
magnetization reversal in an EB-field. At $H_{c1}$ the
magnetization reversal takes place via nucleation and domain wall
movement and leaves the interface in a (partial) magnetic domain
state. The spins in the interfacial domains are strongly coupled
to the antiferromagnet and even in saturation they are not aligned
along the external field direction. Those interface spins give
rise to the strong off-specular diffuse spin-flip scattering. Upon
returning from saturation in negative fields we notice an increase
of the diffuse spin flip scattering, which suggests that a torque
is acting on the ferromagnetic spins trying to reverse them back
to the direction of the cooling field long before the
magnetization reversal actually takes place. It seems that those
interfacial magnetic domains are the seeds of the second
magnetization reversal at $H_{c2}$, which proceeds  via
domain rotation. The domain rotation at $H_{c2}$ is characterized
by a large specular spin-flip scattering (panel c of Fig. \ref{nhl1}) and off-specular
diffuse scattering (not shown). Therefore the second and all other reversals
are dominated  by domain rotation processes.

The behavior of the interface layer as revealed by PNR is in good
qualitative agreement with the observation of uncompensated spins
at the CoO/Co interface by Gruyters and Riegel \cite{gruy} and
with the model put forward by Mauri and co-workers \cite{maur}.
The latter authors suggest the formation of a domain wall parallel
to the interface during reversal of the FM orientation, which can
effectively lower the interfacial energy cost of reversing the FM
layer  without removing the condition of strong interfacial AF/F
coupling. The interface may be considered as a diluted
antiferromagnet. This should be distinguished from the work of
Miltenyi et al.\cite{milt}, who purposely keep the defects away
from the interface.  It would be desirable to image the
interfacial domains by real space methods such as done recently
via polarized electron emission microscopy \cite{ohld}. This
methods can, however, only be applied in remanence. Ferromagnetic
domains as they develop at the interface and in high fields can
presently only be followed via polarized neutron reflectivity.

\section{Conclusions}
In summary, using a wedge type sample with a Co thickness gradient
we have determined the coercive fields and the exchange bias field
after field cooling for different Co layer thicknesses. The
exchange bias field at low temperature follows the $1 / d_{FM}$
behavior. In addition, the temperature dependence of the exchange
bias field exhibits a small positive value just below the blocking
temperature, which is most likely due to a superexchange mechanism
with antiferromagnetic exchange coupling at the AF/F interface.
With polarized neutron reflectivity measurements we have shown
that the first magnetization reversal at $H_{c1}$ is dominated by
nucleation and domain wall motion, while the second magnetization
reversal at $H_{c2}$ can be described by magnetization rotation.
All subsequent reversals are similar to the second, yielding a
symmetric hysteresis, which, however, remains shifted. Furthermore
diffuse spin-flip scattering reveals the formation of magnetic
domains at the interface, induced by the first magnetization
reversal. We have shown that these magnetic domains are preserved
even in saturation and that they serve as seeds for the second
magnetization reversal. Overall, our data reveal clear differences
between the coercive fields of the very first reversal $H_{c1}$
and the second reversal at $H_{c2}$. The mechanisms of the
magnetization reversal (domain wall motion versus domain rotation)
and their temperature dependencies are clearly distinct.

\section{Appendix}

 For the sake of clarity we shall demonstrate
the formulae \ref{SA2} and \ref{SF1} for a simple case. Let us
consider a semi-infinite magnetic media for which we shall obtain
the analytical formulae of reflectivities. We start from the
general formalism presented in Ref.~\cite{radu2}. Using the
expression $\bm{\hat\mu}=\mu \bm{ \hat\sigma}$ and the well known
properties of the Pauli operator $( \bm{ \hat\sigma})$ one can
write the expression of the wave vector  operator as:
$$
\hat k_i^+ = \frac{ k_i^+ + k_i^-}{2}+(\bm{ \hat\sigma
\bm{e_i}})\frac{k_i^+-k_i^-}{2},
$$
where $k_i^\pm=\sqrt{k_z^2-u_i\mp |\bm{\mu B_i}|
\frac{2m}{\hbar^2}}$ are the eigenvalues of the operator $ \hat
k_i^+$ and $ \bm{e_i}=\bm{B_i}/B_i$ is a unit vector which defines
the direction of the magnetic induction in the $i^{\it th}$ media,
$u_i=4\pi N_i b_i^{coh}$ is the nuclear scattering length density
of the i-the layer, where $N$ is the nuclear density and $b^{coh}$
is the coherent scattering length.The matrix representation of
$\hat R_i=\bm{ \hat\sigma e_i}$ reads:

$$
R=\left(\matrix{\cos(\theta))&\sin(\theta) e^{-i\phi}\\ \cr
\sin(\theta)e^{i \phi}&-\cos(\theta) \cr}\right),
$$
where $\theta$ and $\phi$ are the polar and the axial angle in the
spherical coordinates with the polar axis being the quantization
axis. Choosing the quantization axis to be parallel to the
polarization axis, the matrix representation of the reflection
operator $\hat\rho_{01}$ becomes:
$$
\rho_{01}=\left(\matrix{r^{++}&r^{-+}\\ \cr r^{+-}&r^{--}
\cr}\right)
$$
with: \onecolumngrid
\begin{equation}
r^{++}=\frac{{k_1^-}\,{k_0^+} - 2\,{k_1^-}\,{k_1^+} +
{k_0^+}\,{k_1^+} -
    {k_0^-}\,\left( {k_1^-} - 2\,{k_0^+} + {k_1^+} \right)  +
    \left( {k_0^-} + {k_0^+} \right) \,\left( {k_1^-} - {k_1^+} \right) \,\cos (\theta)}{
     {k_1^-}\,{k_0^+} + 2\,{k_1^-}\,{k_1^+} + {k_0^+}\,{k_1^+} +
    {k_0^-}\,\left( {k_1^-} + 2\,{k_0^+} + {k_1^+} \right)  -
    \left( {k_0^-} - {k_0^+} \right) \,\left( {k_1^-} - {k_1^+} \right) \,\cos (\theta)}
\end{equation}

\begin{equation}
r^{--}=-\frac{{k_1^-}\,{k_0^+} + 2\,{k_1^-}\,{k_1^+} +
{k_0^+}\,{k_1^+} -
      {k_0^-}\,\left( {k_1^-} + 2\,{k_0^+} + {k_1^+} \right)  +
      \left( {k_0^-} + {k_0^+} \right) \,\left( {k_1^-} - {k_1^+} \right) \,\cos (\theta)}
      {{k_1^-}\,{k_0^+} + 2\,{k_1^-}\,{k_1^+} + {k_0^+}\,{k_1^+} +
      {k_0^-}\,\left( {k_1^-} + 2\,{k_0^+} + {k_1^+} \right)  -
      \left( {k_0^-} - {k_0^+} \right) \,\left( {k_1^-} - {k_1^+} \right) \,\cos (\theta)}
\end{equation}

\begin{equation}
r^{+-}=\frac{2\,{k_0^+}\,\left( {k_1^-} - {k_1^+} \right) \,e^{i
\,\phi}\,\sin (\theta)}
  {{k_1^-}\,{k_0^+} + 2\,{k_1^-}\,{k_1^+} + {k_0^+}\,{k_1^+} +
    {k_0^-}\,\left( {k_1^-} + 2\,{k_0^+} + {k_1^+} \right)  -
    \left( {k_0^-} - {k_0^+} \right) \,\left( {k_1^-} - {k_1^+} \right) \,\cos (\theta)}
\end{equation}

\begin{equation}
r^{-+}=\frac{2\,{k_0^-}\,\left( {k_1^-} - {k_1^+} \right) \,e^{-i
\,\phi}\,\sin (\theta)}
  {{k_1^-}\,{k_0^+} + 2\,{k_1^-}\,{k_1^+} +
      {k_0^+}\,{k_1^+} + {k_0^-}\,\left( {k_1^-} + 2\,{k_0^+} + {k_1^+} \right)  -
      \left( {k_0^-} - {k_0^+} \right) \,\left( {k_1^-} - {k_1^+} \right) \,\cos (\theta)}
\end{equation}
\twocolumngrid
 It is easy to check that the expressions above are
the same as ones obtained by
 Pleshanov ~\cite{plesh94}.

In order to understand how the reflectivities depend on the
magnetization  reversal in an ferromagnetic media let us consider
the case when $k_z=u$. Choosing just a single value for the
incident wave vector and sweeping the applied magnetic  field is a
key ingredient for measuring a complete hysteresis loop via PNR.
We neglect the neutron absorption into the sample and consider the
case when the applied magnetic field is much smaller then the
internal fields into the ferromagnetic media. With these
assumptions and for $B_x\ge0$ the reflectivities
become:\onecolumngrid
$$
R^{++}=|r^{++}|^2=\frac{1 + {{V_{mn}}}^2 +
2\,{\sqrt{{V_{mn}}}}\,\left( 1 + {V_{mn}} \right) \,\cos (\theta)
+
     2\,{V_{mn}}\,{\cos (\theta)}^2}{{\left( 1 + {\sqrt{{V_{mn}}}} \right) }^2\,
     \left( 1 + {V_{mn}} \right) }
$$
$$
R^{--}==|r^{++}|^2=\frac{1 + {{V_{mn}}}^2 -
2\,{\sqrt{{V_{mn}}}}\,\left( 1 + {V_{mn}} \right) \,\cos (\theta)
+
     2\,{V_{mn}}\,{\cos (\theta)}^2}{{\left( 1 + {\sqrt{{V_{mn}}}} \right) }^2\,
     \left( 1 + {V_{mn}} \right) }
$$
$$
R^{+-}=R^{-+}==|r^{\pm \mp}|^2=\frac{2\,{V_{mn}}\,{\sin
(\theta)}^2}
   {{\left( 1 + {\sqrt{{V_{mn}}}} \right) }^2\,\left( 1 + {V_{mn}} \right) },
$$
\twocolumngrid
where $V_{mn}=Vm/V_n=|\mu B|/(\frac{\hbar^2}{2m} 4 \pi N b_{coh})$ is the
ratio of the magnetic to the nuclear potential. The conclusion
from the formulae above are that at the critical scattering vector
for non-polarized
neutron, the reflectivities:\\
- depend on the absolute value of the magnetization induction;\\
- depend on the relative orientation between the polarization axis
and the direction of the magnetic
induction into the media;\\
-do not depend on the polar angle $\phi$.\\
We consider below two cases: one when the magnetization reversal
proceeds by rotation and another case where the magnetization
changes through domain wall movement.

\subsubsection{Coherent rotation of the  magnetization }

Let us consider that the magnetic moments will rotate during the
magnetization reversal in such a way that $B=B_{sat}$ with the
components $B_y=B \cos (\theta)$ and  $B_x=B \sin (\theta)$. The
easiest way to imagine such a rotation is to rotate the sample
instead of rotating the magnetic moment, keeping the remanent
magnetization equal to the  saturation magnetization
($M_{rem}/M_{sat}=1$).
With this assumption the  spin asymmetry $SA(k_z, \theta, |{\bf
B}|)$ becomes:\onecolumngrid
$$SA(k_z, \theta, {\bf B})=(R^+-R^-)/(R^++R^-)=\frac{2\,{\sqrt{{V_{mn}}}}\,\cos (\theta)}{1 + {V_{mn}}}=\frac{B_x}{{ |{\bf B}|}}\frac{2\,{\sqrt{{V_{mn}}}}}{1 + {V_{mn}}} $$
$$SA(k_z, \theta, {\bf B})=SA(k_z, 0, |{\bf B}|)\frac{B_x}{|{\bf B}|}$$
\twocolumngrid

The spin flip reflectivity is:\onecolumngrid
$$R^{+-}=\frac{2\,{V_{mn}}\,{\sin (\theta)}^2}{{\left( 1 + {\sqrt{{V_{mn}}}} \right) }^2\,\left( 1 + {V_{mn}} \right) }=B^2_y \frac{1}{B^2_{sat}} \frac{2 V_{mn}}{( 1 + \sqrt{V_{mn}})^2 ( 1 + V_{mn} ) }$$
$$R^{+-}=B^2_y  f(B_{sat},V_{mn})$$
\twocolumngrid
Measurements of $R^+$ and $R^-$ at any $k_z$ value
as a function of the applied magnetic field provide the same
hysteresis loop as obtained by MOKE or SQUID measurements.
Moreover, fitting the non-spin flip reflectivity curves taken in
saturation (R$^+$ or R$^-$ or both) one can accurately evaluate
the absolute magnetic moment. In comparison, for MOKE one would
have to make a calibration for the Kerr angles and for SQUID one
needs to determine the volume of the magnetic layer.

\subsubsection{Domain Wall Movement}

Now we discuss a magnetization reversal by domain wall movement.
For simplicity we consider a situation where a magnetic film
consists of ferromagnetic domains with the following
configuration: a fraction of $n$ domains have their spins oriented
parallel to the neutron polarization and a fraction of $1-n$
domains have the spin oriented antiparallel. The measured
reflectivities would not be simply $R^+$ or $R^-$ but given by
$I^+=n R^+ +(1-n)R^-$ and $I^-= (1-n)R^+ +n R^-$. The measured
spin asymmetry becomes:\onecolumngrid

$$SA(k_z,\theta, {\bf |B|})=\frac{I^+-I^-}{I^++I^-}
=(2n-1) (\frac{R^+-R^-}{R^++R^-})=B_x
\frac{1}{B_{sat}}\frac{\sqrt{V_{mn}}}{1 + V_{mn}},$$
\twocolumngrid
where $B_x=(2 n-1){\bf |B|}$ with ${\bf |B|}$ being
the magnetic induction into the magnetic domains. Thus, the shape
of the hysteresis loop is completely defined. The spin flip
scattering will result only from domain walls.

In a more general case when for instance the magnetic domains are
rotated with respect to the neutron polarization axis, there will
also be spin flip scattering. For this case the SA expression will
still be valid and, in addition, from the spin-flip scattering
their angle relative to the neutron polarization axis can be
measured.

We mention that the formulae above are only correct in the absence
of off-specular scattering. The values of  SA and the spin-flip
reflectivity will be affected by the loss of the specular signal.
This can be seen when comparing SQUID or MOKE based magnetic
hysteresis loops to the one extracted from the SA. Regions where
the loops do not overlap indicate the presence of magnetic
domains. This can be clearly seen in Fig. \ref{nhl1}(a).

A different approach for evaluating the average magnetization
vectors inside of a film in a domain state was taken by  Lee et
al. ~\cite{lee}. Instead of measuring the magnetization reversal
at one specific wave vector, complete reflectivity curves are
recorded and the average magnetization as well as the mean square
dispersion of the domains as function of $k_z$ is evaluated. They
use similar reflectivity formulae as shown above, but do not
derive magnetization loops. For a quantitative comparison of
hysteresis loops the method presented here is faster and more
effective.

\acknowledgments

We gratefully acknowledge support through the
Sonderforschungsbereiche 491 "Magnetische Heteroschichten:
Struktur und elektronischer Transport" of the Deutsche
Forschungsgemeinschaft. The neutron scattering experiments were
performed at the ADAM reflectometer of the ILL, which is supported
by the BMBF grant No.~03ZAE8BO. Special thanks are to Erik A.
Verduijn for critical reading of the manuscript.



\end{document}